\documentclass[12pt]{article}
\usepackage{latexsym,epsfig}

\newcommand{\tr}{{\rm\;tr}}
\newcommand{\Tr}{{\rm\;Tr}}
\newcommand{\comm}[2]{\left[#1,#2\right]}
\newcommand{\DD}{{\cal D}}
\newcommand{\const}{{\rm const}}
\renewcommand{\d}{{\rm d}}
\newcommand{\e}{{\rm e}}
\newcommand{\re}{\Re{\rm e}}
\newcommand{\id}{{\rm 1\!\!\!\;l}}
\renewcommand{\i}{{\rm i}}
\newcommand{\binomial}[2]{\left(\begin{array}{c}#1\\#2\end{array}\right)}
\newcommand{\bibit}{\it}
\newcommand{\bibbf}{\bf}

\begin{document}

\noindent DESY 96--154\hfill ISSN 0418-9833\\
December 1996\hfill hep-th/9701124

\renewcommand{\thefootnote}{\fnsymbol{footnote}}

\vspace{2cm}
\begin{center}
{\LARGE Basis Invariants in Non--Abelian Gauge Theories}

\vspace{1cm}
{\large Uwe M\"uller\footnote[1]{E--mail address:
    {\tt umueller@ifh.de}}}\\
  {\it Deutsches Elektronen-Synchrotron DESY,}\\
  {\it Institut f\"ur Hochenergiephysik IfH Zeuthen}\\[0.1ex]
  {\it Platanenallee 6, D--15738 Zeuthen, Germany}
\end{center}
\vspace{2cm}

\setcounter{footnote}{0}
\renewcommand{\thefootnote}{\arabic{footnote}}

\begin{abstract}
  A basis of Lorentz and gauge-invariant monomials in non-Abelian
  gauge theories with matter is described, applicable for the inverse
  mass expansion of effective actions. An algorithm to convert a 
  arbitrarily given invariant expression into a linear combination of
  the basis elements is presented. The linear independence of the basis
  invariants is proven.
\end{abstract}

\section{Introduction}

One-loop effective actions \cite{effac} are space-time
integrals of Lorentz and gauge-invariant expressions. They are closely
related to traces of heat kernels, known as Schwinger-DeWitt
\cite{SDW}, Gilkey-Seeley \cite{GS}, or Hadamard \cite{H}
coefficients. In flat space-time
gauge theory, these coefficients are polynomials consisting of a
matrix potential, the gauge-field strength tensor, and gauge-covariant
derivatives. There are many equivalent forms for these coefficients
essentially due to the Bianchi identity and the product rule for
covariant derivatives. Furthermore, the physically interesting
functional trace allows one to cyclically exchange matrix factors and
to integrate by parts.

Increasing use of computer algebra \cite{belkov etal} and new
methods to calculate effective actions \cite{worldline formalism,
  avramidi} extend our knowledge of heat kernel coefficients 
to higher and higher order. To manage the increasing number of terms
in them and compare results obtained by different methods, a minimal
set of invariants is needed in terms of which all results can be
expressed. In addition, an algorithm to expand a given gauge-invariant
Lorentz scalar into this minimal set should be provided.

This problem was mainly considered in general relativity where the
tensor polynomials consist of the Riemann tensor, the metric 
tensor, and covariant derivatives. This is more complicated than 
flat space-time gauge theory due to the symmetry properties of
the Riemann tensor. However, Fulling et al.\ \cite{FKWC} used 
group representation methods to determine the numbers of independent
monomials, so that an appropriate subset of all monomials can be
chosen to be a basis. Nevertheless, this subset has to be
chosen by hand because there is no known general construction
principle.

For gauge theory in flat space-time, van de Ven \cite{van de Ven}
constructed a basis up to mass dimension ten (fifth order of the
inverse mass expansion). But he also chose the basis elements by hand
and did not present a general construction principle.

For the expansion of effective actions of matter, gauge fields, and
gravity in terms of
Barvinsky-Vilkovisky form factors \cite{bv}, a basis set of non-local
invariants up to third order in the curvature\footnote{In this context
  the gauge-field strength tensor and the matrix potential are
  considered to be the curvatures of the gauge field and the matter
  fields, respectively.}
was defined \cite{BGV}.

The present article, extending our work in Ref.\ \cite{proceedings},
analyses the formal structure of Lorentz and gauge-invariant monomials
in non-Abelian gauge theories with matter in flat 
space-time. Step by step, the operations applicable to these
invariants are used to bring them into a form consisting of
special invariant monomials. These are proven to be linearly 
independent. Thus, a basis set of invariants is specified, and
simultaneously, a procedure to expand a given Lorentz and
gauge-invariant expression in terms of the basis is obtained.

Section 2 introduces notations,
manipulations applicable to invariants, and a
graphical representation of invariants, which is needed for the proof of
linear independence.  Section 3 describes the algorithm to expand a
given gauge-invariant Lorentz scalar in terms 
of special gauge-invariant monomials. The properties of these
monomials are summarized in Section 4. They are used to count the
special monomials up to mass dimension 16. Section 5 shows that this
set of monomials is linearly independent and therefore a
basis. We summarize in Section 6. Appendix A describes
manipulations with fully symmetrized covariant derivatives. The numbers
of independent invariants, subdivided by the number of matrix
potentials and field-strength tensors, are given in Appendix B. Also the
basis invariants up to mass dimension eight are listed there
explicitly. Appendix C describes a construction needed for the proof
of linear independence.

\section{Notations and representation by graphs}

We consider a gauged scalar field theory described by the massive
complex field $\phi^a$ and the hermitian matrix valued gauge
field $A_\mu^{ab}$. The gauge-covariant derivative in the fundamental
representation is ${\cal D}_\mu^{ab}=\delta^{ab}\partial_\mu-\i
A_\mu^{ab}$. The coupling constant is absorbed into the gauge
field. The matter field contribution to the effective one-loop action
is given by
\begin{equation}\label{eff. action}
  \Gamma^{(1)}[\varphi,A]=
  \Tr\ln\left(-\DD^2+m^2+V\right)=
  -\int_0^\infty\frac{\d T}{T}\e^{-Tm^2}
  \Tr\exp\left(-T\left(-\DD^2+V\right)\right)
\end{equation}
The integral in Equation (\ref{eff. action}) is to be understood
regularized at $T=0$ in some way, e.g.\ by dimensional regularization
\cite{dimreg}, by the zeta function procedure \cite{zetareg}
or by a cut-off. The trace of the heat
kernel can be evaluated by various methods \cite{worldline formalism,
  avramidi, methods} and
expanded in gauge-invariant terms \cite{belkov etal, worldline
  formalism, avramidi, van de Ven, methods, results}
\begin{equation}\label{generic result}
  \Tr\exp\left(-T\left(-\DD^2+V\right)\right)=
  \int\d^dx\tr\sum_iC_i{T^{\frac{\mu_i-d}{2}}}I_i\left(F,V\right).
\end{equation}
The matrix potential $V$ is hermitian and originates from the
self-interaction of the
matter field. $C_i$ are in general
complex dimensionless numbers. The space-time dimension is
defined to be $d$. $I_i\left(F,V\right)$ are matrix valued
Lorentz scalars which are composed of the matrix potential $V$, the
field strength tensor
\begin{equation}
F_{\mu\nu}^{ab}=\i\comm{\DD_\mu}{\DD_\nu}^{ab}=
  \partial_\mu A_\nu^{ab}-\partial_\nu A_\mu^{ab}-
  \i\comm{A_\mu}{A_\nu}^{ab},
\end{equation}
and the adjoint gauge-covariant derivative
\begin{equation}
D_\mu=\comm{\DD_\mu}{\,}=\partial_\mu-\i \comm{A_\mu}{\,}.
\end{equation} 
$D_\mu$ acts on $V$ and on $F_{\mu\nu}$.
$\mu_i$ is the mass dimension of the Lorentz scalar
$I_i\left(F,V\right)$ according to the mass dimensions of
its constituents $[V]=2$, $[F_{\mu\nu}]=2$, and $[D_\mu]=[\DD_\mu]=1$.
The first few terms of the expansion (\ref{generic result}) are
\cite{results} 
\begin{eqnarray}
  \Tr\exp\left(-T\left(-\DD^2+V\right)\right)=
  \left(4\pi T\right)^{-\frac{d}{2}}
  \int\d^dx\tr\left[\id-TV+\frac{T^2}{2}
  \left(V^2+\frac{1}{6}F_{\mu\nu}F_{\mu\nu}\right)+ 
  \ldots\right].
\end{eqnarray}
The form (\ref{generic result}) is not unique due to the possible
manipulations of Table \ref{manip}. In this article, we call the
possibility of cyclic matrix permutations under the trace {\em cyclic
  invariance}, and that of mirror transformations {\em reflection
  invariance}. The latter stems from the possibility of matrix
transpositions under the trace. The scope of the present article is to
fix all manipulation possibilities and to obtain a basis for the
gauge-invariant\footnote{The objects $I_i(F,V)$ are Lorentz invariant
  but, in the strict sense, not gauge invariant. However, here we
  understand $I_i(F,V)$ always to stand inside a trace, which results
  in gauge-invariant quantities, and call them loosely invariants,
  too.} Lorentz scalars $I_i(F,V)$.

\begin{table}[htb]
  \caption{\footnotesize The possible manipulations applicable to
    Lorentz scalars $I_i(F,V)$ and the corresponding equalities that
    are used. Integration by parts assumes the background fields
    decrease sufficiently quickly at infinity. In the last line, $X$,
    $Y$, and $Z$ are assumed to be ``simple factors'' (defined in the
    text). Further relations between invariants such as equations of
    motion or additional symmetries are not taken into account.} 
  \label{manip}\vspace{2mm}
  \begin{tabular}{|l|l|}
    \hline
    Manipulation&Used equality\\
    \hline
    Product rule&$D_\mu(XY)=D_\mu XY+XD_\mu Y$\\
    Cyclic matrix permutations&$\tr(XYZ)=\tr(YZX)$\\
    Integration by parts&$ \int\d x\tr\left(D_\mu X_\mu Y\right)=
    -\int\d x\tr\left(X_\mu D_\mu Y\right)$\\
    Bianchi identity&$D_\mu F_{\kappa\lambda}=
    D_\kappa F_{\mu\lambda}+D_\lambda F_{\kappa\mu}$\\
    Antisymmetry of $F_{\mu\nu}$&$F_{\mu\nu}=-F_{\nu\mu}$\\
    Mirror transformation&$\overline{\tr\left(XYZ\ldots\right)}=
    \tr\left(\ldots Z^\dagger Y^\dagger X^\dagger\right)=
    \tr\left(\ldots ZYX\right)$\\
    \hline
  \end{tabular}
\end{table}

We introduce some further terminology. $V$, $F_{\kappa\lambda}$,
$D_{\mu_1}D_{\mu_2}\ldots D_{\mu_n}V$, or 
$D_{\mu_1}D_{\mu_2}\ldots D_{\mu_n}F_{\kappa\lambda}$ ($n\ge
1$) are {\em simple factors}. Containing $V$ they are called {\em $V$-factors}, otherwise {\em
  $F$-factors}. {\em Simple invariants} or {\em invariant monomials}
are products of simple factors.

Due to the hermiticity of $A_\mu$ simple factors are always hermitian
\begin{equation}\label{hermiticity}
  V^\dagger=V,\quad\quad
  F_{\mu\nu}^\dagger=F_{\mu\nu},\quad\quad
  X^\dagger=X\quad\Longrightarrow\quad
  \left(D_\mu X\right)^\dagger=D_\mu X.
\end{equation}
This property provides the basis for the reflection invariance of simple
invariants. 

Applying the product rule for covariant derivatives,
the invariants $I_i(F,V)$ of Equation (\ref{generic result})
may always be expressed by sums of simple invariants. Therefore
we assume in the following all invariants $I_i(F,V)$ to be simple.

The last manipulation of Table \ref{manip} is called mirror
transformation because it inverts the ordering of the factors $X$,
$Y$, and $Z$. It relates an invariant with the {\em complex
conjugate} of another one. Generically, a quantity and its complex
conjugate are linearly independent, and the mirror transformation
cannot be used to substitute an invariant by another one. But in
special situations this transformation {\em is} useful.
Assuming the effective action (\ref{generic result})
is real, then --- expanded in an appropriate basis --- every simple
invariant $I_1=\tr\left(XYZ\ldots\right)$ and its mirror image
$I_2=\tr\left(ZYX\ldots\right)=\overline{I_1}$ have coefficients $C_1$ and
$C_2$ complex conjugate to each other, so that the sum of them can be
written as $C_1 I_1+C_2 I_2=2\re\left(C_1 I_1\right)$. Obviously $I_2$
was replaced by $I_1$ but the price paid is the introduction of the
projection operator $\re$. Another special situation is a
real field $\phi^a$ and a real representation of the gauge group.
$V$ is then real and symmetric, whereas $A_\mu$ and $F_{\mu\nu}$ are
imaginary and antisymmetric (thus still hermitian).
The simple invariants and its complex conjugates are then equal up to
a factor of $(-1)^f$ (where $f$ is the number of $F$-factors in the
corresponding invariant).

The ordering of derivatives within a factor can be changed by continued
application of
\begin{equation}\label{commute}
  D_\mu D_\nu X=D_\nu D_\mu X-\i\comm{F_{\mu\nu}}{X}.
\end{equation}
This produces additional terms with more $F$-factors and fewer derivatives.
After the commutation of some inner derivatives the product rule
has to be applied again. In this way every ordering of
derivatives can be achieved.

The indices in a simple invariant can be contracted
between different factors and within the same factor. Let us consider
the latter case, we call it {\em self-contraction}. At least
one index must belong to a derivative, 
otherwise it would be the contraction $F_{\mu\mu}$ which forces the
whole invariant to vanish.
After commuting this derivative to the outside of the factor,
integration by parts can be applied. In the resulting expression one
self-contraction has been eliminated.
Repeating this, we arrive at an expression where all invariants
$I_i(F,V)$ have no self-con\-trac\-tions. From now on we assume our
gauge- and Lorentz-invariant expression (\ref{generic result}) to have
this form.

\begin{figure}[htb]
  \begin{center}
    (a)$\quad$\epsfig{file=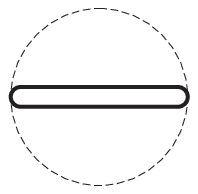,width=1.5cm}\hfill
    (b)$\quad$\epsfig{file=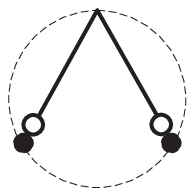,width=1.5cm}\hfill
    (c)$\quad$\epsfig{file=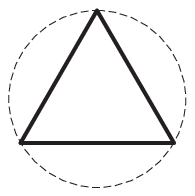,width=1.5cm}\hfill
    (d)$\quad$\epsfig{file=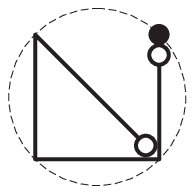,width=1.5cm}
  \end{center}
  \caption{\footnotesize
    Examples for the graphical representation of terms: (a)
    $\tr(F_{\mu\nu}F_{\mu\nu})$, (b) $\tr(F_{\mu\nu}D_\mu VD_\nu V)$, (c)
    $\tr(F_{\mu\nu}F_{\kappa\mu}F_{\nu\kappa})$, (d)
    $\tr(F_{\mu\nu}F_{\kappa\mu}D_\nu F_{\lambda\kappa}D_\lambda V)$.}
    \label{f1}
\end{figure}

Now we are able to introduce the graphical representation of simple
invariants. The factors are represented by regions located on a circle
(cf.\ Tab.~\ref{t1} and Fig.~\ref{f1}). The circle takes into account
the cyclic invariance of the trace. Reading the invariant from
left to right corresponds to reading the graph counterclockwise along
the circle.
\begin{table}[htb]
  \caption{\footnotesize
    Graphical elements for the representation of simple
    invariants}\label{t1}
  \vspace{1ex}\footnotesize
  \begin{center}
    \begin{tabular}{cc}
      \hline
      Part of a term & Graphical representation\\
      \hline
      $V$ & $\bullet$\\[1mm]
      $F_{\mu\nu}$ &
      \unitlength0.6pt
      \begin{picture}(40,0)
        \put(0,5){\line(5,1){30}}
        \put(0,5){\line(5,-1){30}}
        \put(32,7){$\mu$}
        \put(32,-5){$\nu$}
      \end{picture}\\[1mm]
      $D_\mu$ & $\circ\hspace{-0.85ex}
      -\!\!\!-\hspace{-10pt}-\!\!\!-\!\!\!-\hspace{-2pt}\mu$\\
      \hline
    \end{tabular}
  \end{center}
\end{table}
Contractions between factors
are represented by straight lines connecting the regions.
Open points at the line ends symbolize derivatives. $V$'s are depicted
by closed points. $F_{\mu\nu}$'s are recognizable by two linked lines
representing the two indices. The representation
of simple invariants by graphs is unambiguous only modulo permutations of
derivatives (\ref{commute}) and index interchanges in
$F_{\mu\nu}$'s. This ambiguity can be eliminated by choosing a 
certain index ordering for the invariants. This will be
done below where the algorithm for expanding a given invariant in basis
invariants is described.

\section{The algorithm}

This section describes an algorithm using the manipulations
of Table \ref{manip} to
transform an arbitrary expression (\ref{generic result}) into a
standard form. Section 5 will show that the invariant monomials
left unchanged by the algorithm are linearly independent, at least if
we exclude mirror transformations from the possible manipulations of
Table  \ref{manip} (including mirror transformations the same is
true at least up to 16 mass dimensions, as we will see
below). Therefore the result of the algorithm, expressed by 
the basis invariants, will always be unique, though the algorithm may
offer at intermediate stages several possibilities to proceed.

We start with an arbitrary Lorentz and gauge-invariant expression of
the form (\ref{generic result}). We treat invariants with different
mass dimension separately, since the manipulations of Table
\ref{manip} do not mix them.
In Section 2 the product rule and 
integration by parts were used to obtain a sum
of simple invariants which have no self-contractions and can be
represented by graphs. {\em This is the
  first step of the algorithm}.

The following steps require commutations of derivatives by Equation
(\ref{commute}), which produce additional invariant monomials
containing more $F$-factors and fewer derivatives. Therefore we
concentrate on the invariant monomials {\em not in their
  final form} and having the {\em fewest $F$-factors} (or alternatively,
the most derivatives, i.e.\ {\em the fewest factors}). Then the
following steps of the algorithm are applied to 
these monomials, transforming them into their final form and falling
out of the consideration. This procedure is iterated until all
monomials are in their final form.

\begin{figure}[htb]
  \begin{center}
  \makebox[0cm][l]{\hspace{0.9cm}\raisebox{1cm}{L}}
  \makebox[0cm][l]{\hspace{1.6cm}\raisebox{2cm}{M}}
  \makebox[0cm][l]{\hspace{2.5cm}\raisebox{1cm}{R}}
  \epsfig{file=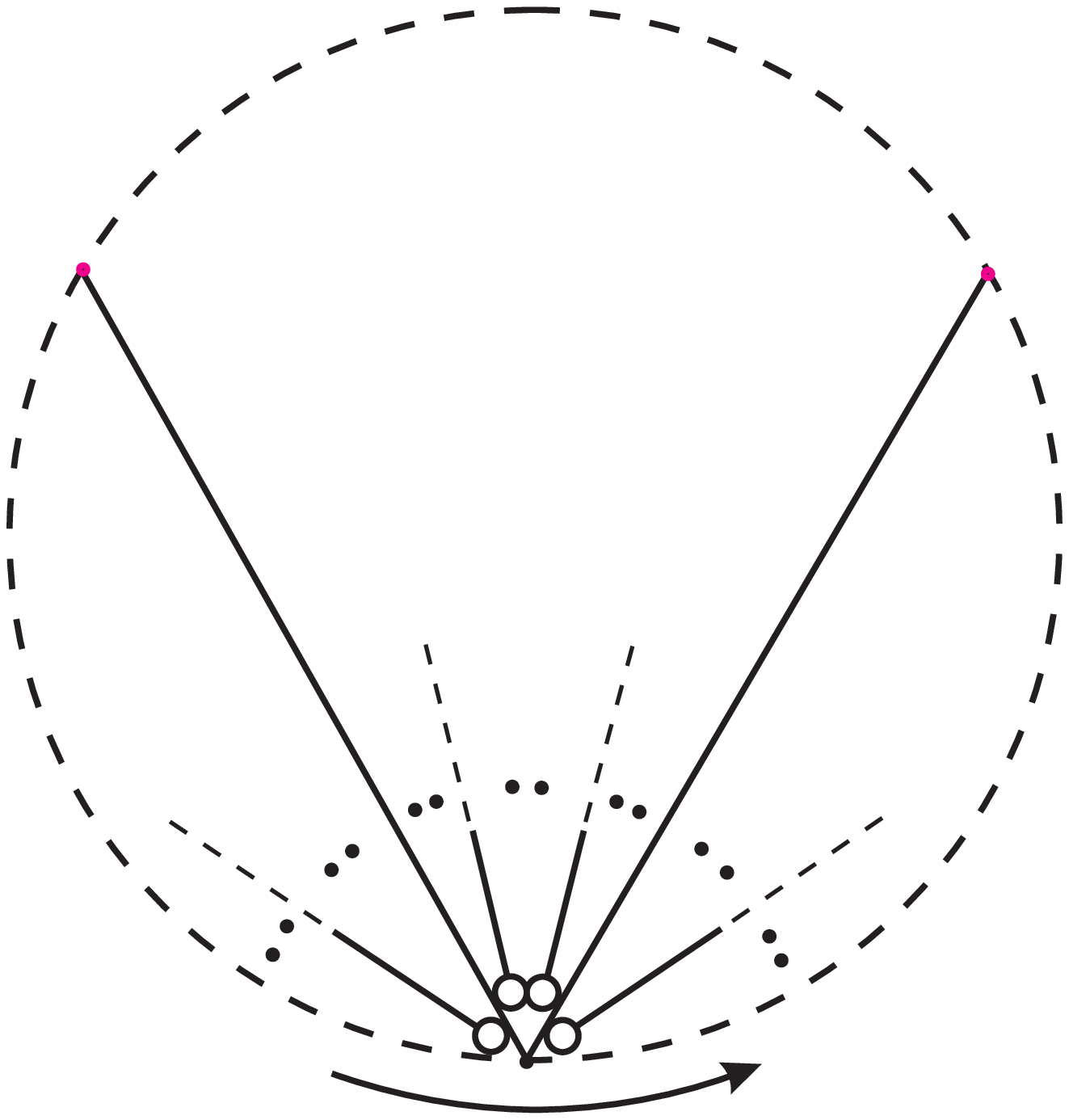, width=3cm}
  \end{center}\vspace{-2ex} 
  
  \caption{\footnotesize
    A certain $F$-factor within a graph. The two $F$-indices
    give rise to three sectors, marked with L, M, and R, with which
    the derivatives are contracted. Following the arrow corresponds
    to reading the invariant monomial from the left to the right.}
  \label{f2}
\end{figure}

The first step inside the iteration is to use the Bianchi
identity. This identity interchanges the index of a derivative 
with the indices of the field strength tensor within an $F$-factor,
without disturbing the other factors.
To apply the Bianchi identity
to an $F$-factor with more than one derivatives we must specify
the derivative which we want to use and commute it with the other
derivatives by Equation (\ref{commute}) so that it becomes the
innermost derivative. Which derivatives do we specify?
We look at the example
\begin{equation}\label{example sectors}
  \tr(
  \underbrace{
  \stackrel{\mbox{L}}{D_\mu}
  \stackrel{\mbox{R}}{D_\nu}
  \stackrel{\mbox{M}}{D_\rho}
  D_\sigma
  F_{\kappa\lambda}}_{\makebox[0cm]{\parbox{2.5cm}{\centering factor
      under\\consideration}}} 
  \underbrace{\ldots X_\nu'\ldots}_{\parbox{1.5cm}{\centering right sector}}
  Y_{\sigma\kappa}
  \underbrace{\ldots X_\rho''\ldots}_{\parbox{1.5cm}
    {\centering middle sector}}Z_\lambda
  \underbrace{\rule{0cm}{1.85ex}
    \ldots X_\mu\ldots}_{\makebox[0cm][l]{\parbox{1.5cm}{\centering
        left sector}}})\quad.
\end{equation}
The indices of $F_{\kappa\lambda}$ are contracted with the factors
$Y_{\sigma\kappa}$ and $Z_\lambda$, which divide the remaining factors
into three, possibly empty, sectors. These are denoted by ``right
sector'', ``middle sector'', and ``left sector'', because
the ``left sector'' is connected with the left-hand side of the
considered factor by cyclic invariance. This is
depicted in Figure \ref{f2}. 

The derivatives of the factor under consideration are called 
left (``L''), right (``R''), and middle (``M'') corresponding to the
sector with which they are contracted. Not all derivatives are left,
right, or middle (e.g.~$D_\sigma$). The Bianchi identity
mixes all three kinds of derivatives. 
Therefore it can be used to
eliminate one kind of derivative in all factors of all monomials. Since
the middle sector is invariant under the mirror transformation (left and
right sectors are interchanged), {\it we apply the Bianchi identity to
  middle derivatives.} Each such application of the Bianchi identity
reduces the number of factors in the corresponding middle
sector. So after a finite number of steps, all middle derivatives are
eliminated.

Next, we convert multiple contractions between factors into a
standard form by 
\begin{eqnarray}\label{multiple1}
  \ldots F_{\mu\nu}\ldots D_\mu D_\nu X\ldots&\Rightarrow&
  -\frac{\i}{2}\ldots F_{\mu\nu}\ldots\comm{F_{\mu\nu}}{X}\ldots\\
  \label{multiple2}
  \ldots F_{\mu\nu}\ldots D_\mu F_{\nu\kappa}\ldots&\Rightarrow&
  \frac{1}{2}\ldots F_{\mu\nu}\ldots D_\kappa F_{\nu\mu}\ldots\\
  \label{multiple3}
  \ldots D_\mu F_{\nu\kappa}\ldots D_\nu F_{\mu\lambda}\ldots&\Rightarrow&
  \ldots D_\mu F_{\nu\kappa}\ldots D_\mu F_{\nu\lambda}\ldots+\nonumber\\
  &&\hspace{1cm}+
  \frac{1}{2}\ldots D_\kappa F_{\nu\mu}\ldots D_\lambda F_{\mu\nu}\ldots.
\end{eqnarray}
The first transformation uses the antisymmetry of
the field strength tensor and the commutation rule (\ref{commute}).
The second transformation relies again on the antisymmetry
together with the Bianchi identity. The third equality results from
the Bianchi identity for one of the factors and the second 
transformation (\ref{multiple2}).

The graphical representation of monomials automatically takes into
account the cyclic invariance if we
regard a certain graph and the same graph but rotated as 
identical. The reflection invariance can be
taken into account by identification of graphs with their mirror
images. In the analytic expressions we pick a
representative of each class of equivalent ``rotated'' and
``reflected'' monomials, for instance the first or last monomial
according to some lexicographical order. 

The last step inside the above mentioned iteration is to arrange
the derivatives and the indices of the field strength tensor in a
definite way. This will be done factor by factor, such as for the
application of the Bianchi identity. Here three prescriptions are
proposed.

{\bf 1)} We consider a certain factor
of a simple invariant. Temporarily suspending the fixing of the cyclic
invariance, the factor is shifted to the left end of this monomial,
as in example (\ref{example sectors}). We call two indices which are
contracted 
with each other {\em partners} for the moment. We then rearrange the
derivatives and indices of the field strength tensor (if present) in
the considered factor so that they copy the ordering of their partner
indices. In example (\ref{example sectors}) $Y_{\sigma\kappa}$ is
located left of $Z_\lambda$, therefore the indices of
$F_{\kappa\lambda}$ have the correct order. The locations of $X_\nu'$,
$Y_{\sigma\kappa}$, 
$X_\rho''$, and $X_\mu$ define the correct order of the
derivatives to be $D_\nu D_\sigma D_\rho D_\mu$. While this
arrangement of indices is independent of the fixing of 
the cyclic invariance due to the temporary shift, it is still
dependent on mirror transformations, which invert the ordering of the
factors. If we choose a different fixing of the reflection
invariance we obtain a slightly different basis which results in
different coefficients in front of the basis invariants. Since there
appears to be no natural way to fix the reflection 
invariance, this prescription for the ordering of the
indices seems to be sensible only if the problem we want to describe
does not allow mirror transformations.

{\bf 2)} The second prescription is relatively simple and has no
problems with the reflection invariance. The derivatives are
symmetrized by forming the sum over all index 
permutations:\footnote{We use here the notation
$D_{\alpha_1\cdots\alpha_k}=D_{\alpha_1}\cdots D_{\alpha_k}$.}
\begin{equation}
  D_{\alpha_1\cdots\alpha_k}X\to
  D_{(\alpha_1\cdots\alpha_k)}X=
  \frac{1}{k!}\sum_{\pi\in S_k}
  D_{\alpha_{\pi(1)}\cdots\alpha_{\pi(k)}}X.
\end{equation}
This can be realized by successive application
of\/\footnote{$(\alpha_1\cdots\alpha_{\lambda-1}|\mu|\alpha_\lambda
\cdots\alpha_k)$ means that $\mu$ is excluded from the
symmetrization.}
\begin{equation}\label{symm recursion}
  D_\mu D_{(\alpha_1\cdots\alpha_k)}X=
  D_{(\mu\alpha_1\cdots\alpha_k)}X
  -\frac{\i}{k+1}\sum_{\lambda=1}^k\binomial{k+1}{\lambda+1}
  \comm{D_{(\alpha_1\cdots\alpha_{\lambda-1}|}F_{\mu|\alpha_\lambda}}
  {D_{\alpha_{\lambda+1}\cdots\alpha_k)}X}.
\end{equation}
This equation is derived in Appendix A.
The indices of the field strength tensor are ordered 
as in the first prescription or somehow different --- this is not so
important because it produces only factors of $(\pm 1)$ and different
conventions of this point can easily be converted. Working with this
prescription, it is more practical to deal with symmetrized derivatives
from the very beginning.\footnote{The calculation of effective
  actions by the 
  world-line formalism \cite{worldline formalism} yields automatically
  fully symmetrized derivatives. Furthermore, the expressions resulting
  from this formalism have no self-contractions, thus the first step of
  the algorithm is already completed from the outset.}
Formulae relevant for this case are given in Appendix A.  

{\bf 3)} The third prescription is a compromise between the first
two. As in the previous paragraph, the ordering of the
indices of the field strength tensor is not so important. 
But instead of the fully symmetrized derivatives of prescription 2) we
form the sum over one ordering of derivatives and the opposite
ordering 
\begin{equation}
  D_{\alpha_1\cdots\alpha_k}X\to
  \frac{1}{2}\left(D_{\alpha_1\alpha_2\cdots\alpha_k}X+
  D_{\alpha_k\alpha_{k-1}\cdots\alpha_1}X\right).
\end{equation}
This is sufficient to ensure reflection invariance of the basis
invariants. The ordering $D_{\alpha_1\alpha_2\cdots\alpha_k}$ of the
derivatives itself is adjusted as in the first prescription.

While the first prescription depends on the
fixing of the reflection invariance, the second and third ones do
not. At least if the invariants of our theory can be mirror
transformed, this strongly suggests the use of one of the latter
prescriptions to give results in a definite form. If reflection
invariance is absent, there is in principle no difference between the
three prescriptions. In all cases, the first prescription seems to be
most appropriate for numerical calculations since the basis invariants
are simple. In the second and third prescription, the basis invariants
are not simple but symmetrized sums of simple invariants. While this
does not affect the algorithm nor the proof of the basis property, it
inflates the number of terms in numerical calculations, especially
with the second prescription. The third prescription seems to be a good
compromise between brevity and definiteness under reflection
invariance.

\section{Properties of the basis invariants}

Pursuing the lines of the above algorithm we can characterize the
basis invariants by the following properties:
\begin{itemize}
  \item The invariants are (symmetrized sums of) products of simple
    factors.
  \item Indices are contracted only between different factors of an
    invariant monomial.
  \item There are no middle derivatives.
  \item In multiple contractions between factors, 
    derivatives are contracted with de\-ri\-va\-ti\-ves and indices of
    $F$'s with indices of $F$'s if this is possible to arrange by the
    transformations (\ref{multiple1} -- \ref{multiple3}). That means,
    index contractions like the ones on the left-hand side of the
    transformations (\ref{multiple1} -- \ref{multiple3}) do not occur
    in the basis invariants.
  \item The order of derivatives and of indices of the $F$'s is given
    by one of the prescriptions of the last section. The main issue
    here is that such a prescription makes the assignment between
    graphs and invariants unambiguous.
\end{itemize}
These properties allow one to count the basis invariants of a certain
mass dimension. In fact, it is more convenient to count the graphs
assigned to the basis invariants. Up to mass dimension 16, this was
performed by a C language computer program with the results shown in
Table \ref{number of invariants} and Appendix B.
\label{homepage}
The program can be obtained soon by accessing the author's home page
on the DESY-IfH web site
(http://www.ifh.de/{$\sim$}umueller/invariants.html).

\begin{table}[htbp]
  \centering
  \caption{\footnotesize The number of invariants {\bf with} and {\em
      without} identification under the mirror transformation. $v$ is
    the number of occurrences of the matrix potential in the
    invariants.}
  \label{number of invariants}
  \vspace{1ex}\footnotesize
  \raggedright
  \begin{tabular}{|c|c||rr||rr|rr|rr|}
    \hline
    &Mass dim.&\multicolumn{2}{c||}{Total}&
    \multicolumn{2}{c|}{$v=0$}&\multicolumn{2}{c|}{1}&
    \multicolumn{2}{c|}{2}\\
    \hline
    1&2&\bf 1&\it 1&\bf 0&\it 0&\bf 1&\it 1&&\\
    2&4&\bf 2&\it 2&\bf 1&\it 1&\bf 0&\it 0&\bf 1&\it 1\\
    3&6&\bf 5&\it 5&\bf 2&\it 2&\bf 1&\it 1&\bf 1&\it 1\\
    4&8&\bf 17&\it 18&\bf 7&\it 7&\bf 4&\it 5&\bf 4&\it 4\\
    5&10&\bf 79&\it 105&\bf 29&\it 36&\bf 24&\it 36&\bf 17&\it 23\\
    6&12&\bf 554&\it 902&\bf 196&\it 300&\bf 184&\it 329&\bf 119&\it 191\\
    7&14&\bf 5283&\it 9749&\bf 1788&\it 3218&\bf 1911&\it 3655&
    \bf 1096&\it 2020\\
    8&16&\bf 65346&\it 127072&\bf 21994&\it 42335&\bf 24252&\it 47844&
    \bf 13333&\it 25861\\
    \hline
  \end{tabular}\\[1ex]
  Table \ref{number of invariants}. (Continued)\\
  \begin{tabular}{|c|c||rr|rr|rr|rr|rr|rr|}
    \hline
    &Mass dim.&\multicolumn{2}{c|}{$v=3$}&\multicolumn{2}{c|}{4}&
    \multicolumn{2}{c|}{5}&\multicolumn{2}{c|}{6}&
    \multicolumn{2}{c|}{7}&\multicolumn{2}{c|}{8}\\
    \hline
    1&2&&&&&&&&&&&&\\
    2&4&&&&&&&&&&&&\\
    3&6&\bf 1&\it 1&&&&&&&&&&\\
    4&8&\bf 1&\it 1&\bf 1&\it 1&&&&&&&&\\
    5&10&\bf 6&\it 7&\bf 2&\it 2&\bf 1&\it 1&&&&&&\\
    6&12&\bf 39&\it 63&\bf 13&\it 16&\bf 2&\it 2&\bf 1&\it 1&&&&\\
    7&14&\bf 370&\it 670&\bf 96&\it 158&\bf 18&\it 24&\bf 3&\it 3&
    \bf 1&\it 1&&\\
    8&16&\bf 4452&\it 8638&\bf 1095&\it 2020&\bf 186&\it 329&
    \bf 30&\it 41&\bf 3&\it 3&\bf 1&\it 1\\
    \hline
  \end{tabular}
\end{table}

\section{Proof of linear independence}
\label{proof}

This section proves that the set of invariant monomials described in
Section 4 is linearly independent. Since every gauge- and
Lorentz-invariant expression of the form (\ref{generic result}) can be
written as a linear combination of this set, as realized by the
algorithm, then it follows that this set of invariants is a basis.

The proof relies on a special background configuration that was
used by van de Ven to compute one-loop counter-terms up to
ten space-time dimensions \cite{van de Ven}. It is given by 
\begin{equation}\label{special AV} 
  A_\mu=\const,\qquad V=-A_\mu A_\mu=\const.
\end{equation} 
This implies 
\begin{equation}\label{special DF} 
  D_\mu X=\comm{\DD_\mu}{X}=-\i\comm{A_\mu}{X},\qquad
  F_{\mu\nu}=-\i\comm{A_\mu}{A_\nu}.
\end{equation} 
Replacing $V$, $F$, and $D$
by the constant field $A$, every invariant monomial $I(F,V)$ has the
form 
\begin{equation}\label{special invariant}
I(F,V)=\sum_k c_k A_{\nu_1}A_{\nu_2}\ldots A_{\nu_{\mu}}= \sum_k c_k
J_k(A). 
\end{equation} 
The mass dimension of $I(F,V)$ is $\mu$, $c_k$ are some complex
numbers. We call the products of $A$-matrices $J_k(A)$
{\em primitive invariants}. Table \ref{manipulations} shows
how the possible manipulations of invariants in a general background
translate into manipulations of primitive invariants.

The primitive invariants can be represented by graphs where the $A$'s
are depicted as points located on a circle and the contractions
between them are straight lines connecting the points. These graphs
were used also in Ref.\ \cite{van de Ven}.
To distinguish them from the ones of section 2 we call them
{\em primitive graphs}.

\begin{table}[htbp]
  \centering
  \caption{\footnotesize
    The manipulation possibilities for invariants. Primitive
    invariants are defined in the text.}
  \label{manipulations}
  \vspace{1ex}\footnotesize
  \begin{tabular}{c@{\hspace{10ex}}c}
    \hline
    General invariants&Primitive invariants\\ 
    \hline 
    Integration by parts&Cyclic matrix exchanges\\
    Cyclic matrix exchanges&Cyclic matrix exchanges\\
    Bianchi identity&Is trivially fulfilled\\
    Mirror transformation&Mirror transformation\\
    \hline
  \end{tabular}
\end{table}

In Ref.\ \cite{van de Ven} it was found that the specialization to
this background configuration does not change 
the number of independent invariants up to ten dimensions. This
generalizes to all mass dimensions in the following proposition.

{\bf Proposition. }{\em
  The number $n_p$ of independent primitive invariants with a certain
  mass dimension is equal to the number $n_b$ of invariants specified
  in Section 4 with the same mass dimension if we do not
  use the mirror transformation. Moreover these numbers are equal to
  the maximal numbers $n_{\rm max}$ and $n_{{\rm max},0}$ of
  independent invariants in general and in the special background
  (\ref{special AV}),
  respectively.}

{\bf Proof.} First, we show that the set of independent primitive
invariants is as large as the set of independent invariants at
the special background. From (\ref{special invariant}) it 
is clear that every invariant at the special background can be
expressed by primitive invariants. The converse can be established by
providing a set of rules to construct an expression of general
invariants being equal at the special background to a given primitive
invariant. These rules are described in Appendix C. Thus we have
$n_p=n_{{\rm max},0}$.

Second, we note that independent invariants in the special
background are also independent in general backgrounds. Thus
we have $n_{{\rm max},0}\le n_{\rm max}$.  
\begin{table}
  \centering
  \caption{\footnotesize
    One-to-one mapping between primitive graphs and graphs
    representing basis invariants. It maps lines to lines.
    For an additional explanation see the text. Some lines in the
    primitive graphs are bent in this table to depict exactly
    which lines they intersect.}
  \label{mapping}
  \newlength{\graphwidth}
  \graphwidth4.5cm
  \vspace{2ex}\footnotesize
  \begin{tabular}{ccc}
    \hline
    \multicolumn{1}{c}{\hspace*{8.5mm}Primitive graphs\hspace*{8.5mm}}
    &&\multicolumn{1}{c}{Graphs representing basis invariants}\\
    \hline
    &&\\
    \parbox{\graphwidth}
    {\epsfig{file=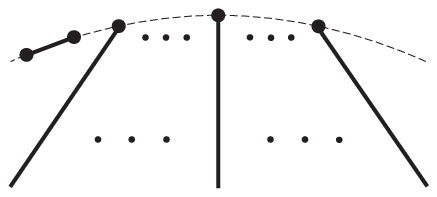,width=\graphwidth}}&
    $\longleftrightarrow$ &
    \parbox{\graphwidth}
    {\epsfig{file=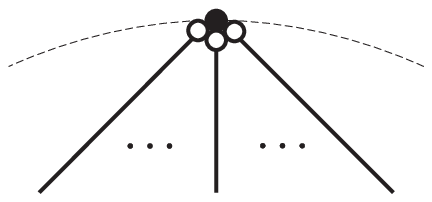,width=\graphwidth}}\\
    &&\\
    \parbox{\graphwidth}
    {\epsfig{file=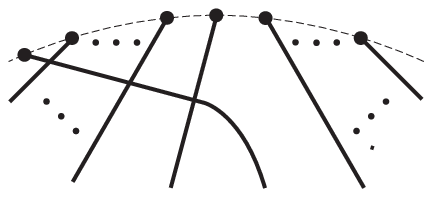,width=\graphwidth}}&
    $\longleftrightarrow$&
    \parbox{\graphwidth}
    {\epsfig{file=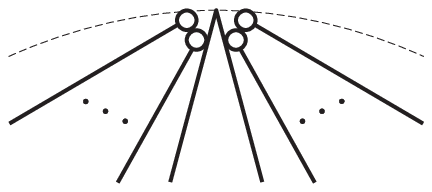,width=\graphwidth}}\\
    &&\\
    \hline
    \multicolumn{3}{c}{Special cases: multiple lines between two factors}\\
    \hline
    &&\\
    \parbox{\graphwidth}
    {\epsfig{file=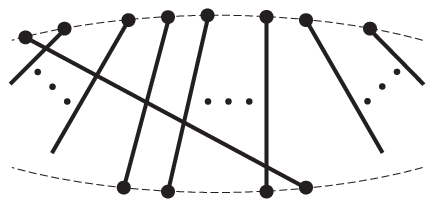,width=\graphwidth}}&
    $\longleftrightarrow$&
    \parbox{\graphwidth}
    {\epsfig{file=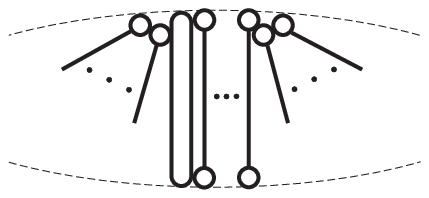,width=\graphwidth}}\\
    &&\\
    \parbox{\graphwidth}
    {\epsfig{file=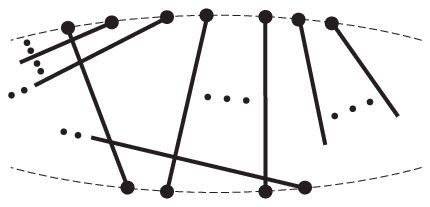,width=\graphwidth}}&
    $\longleftrightarrow$&
    \parbox{\graphwidth}
    {\epsfig{file=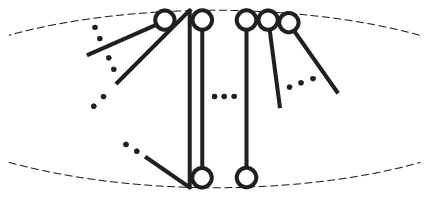,width=\graphwidth}}\\
    &&\\
    \parbox{\graphwidth}
    {\epsfig{file=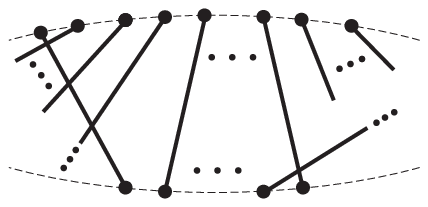,width=\graphwidth}}&
    $\longleftrightarrow$&
    \parbox{\graphwidth}
    {\epsfig{file=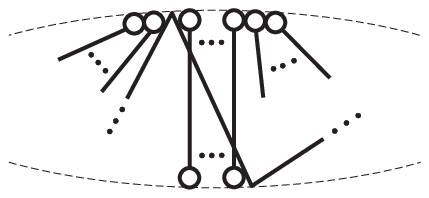,width=\graphwidth}}\\
    &&\\
    \parbox{\graphwidth}
    {\epsfig{file=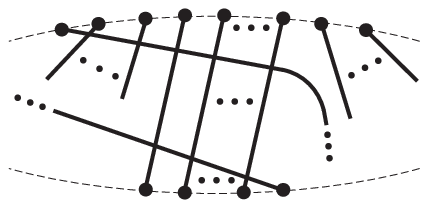,width=\graphwidth}}&
    $\longleftrightarrow$&
    \parbox{\graphwidth}
    {\epsfig{file=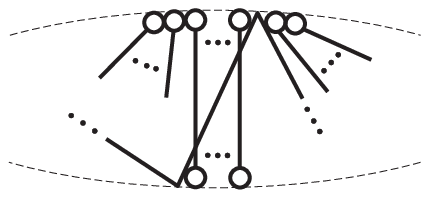,width=\graphwidth}}\\
    &&\\
    \hline
    \end{tabular}
\end{table}

Third, we show that the number of primitive invariants $n_p$ equals
the number $n_b$ of invariants achieved by our algorithm. This
is done by providing a one-to-one mapping between the corresponding
graphs representing the invariants. The mapping is given in Table
\ref{mapping}\label{expl}. Lines are mapped to lines. The division of
primitive graphs into 
factors is determined by the intersection behaviour of lines. Going
counterclockwise around a graph, the last two points of a factor are
connected ($V$-factor) or have intersecting lines ($F$-factor). All
points before the last point and belonging to the same factor have
non-intersecting lines. This determines unambiguously which points 
belong together forming a factor in a primitive graph.

The last step in this proof is to note that by our algorithm every
invariant in general backgrounds can be expressed by our
invariants. This means $n_{\rm max}\le n_b$. Looking at all relations
so far derived 
\begin{equation}\label{proof finish}
  n_p=n_{{\rm max},0}\le n_{\rm max}\le n_b=n_p
\end{equation}
we conclude from the equality of the first and the last number the
equality of all numbers involved. $\Box$

We observe that the mapping (Table \ref{mapping}) does
not preserve the symmetry of the mirror transformation. Therefore two
graphs equivalent by the mirror transformation may be mapped to two
inequivalent graphs. Thus this mapping is not one-to-one in this case, so
the proof fails. The rest of
the proof holds also with the mirror transformation. To show
the linear independence of the set of invariants in this case, it is therefore
sufficient to establish $n_b=n_p$. Up to a fixed order of mass
dimension this can be done simply by counting the invariants. Table
\ref{number of invariants} gives the number of primitive invariants
up to mass dimension 16. Up to this dimension, the equality of $n_b$
and $n_p$ was checked to hold also in the presence of the mirror
transformation.

\section{Summary}

We obtained a prescription for the construction of a set of invariants
in non-Abelian 
gauge theories. A reduction algorithm was presented to transform a
given Lorentz scalar by integrations by parts, cyclic matrix
exchanges, and Bianchi identities into a linear combination of this
set of invariants. It proves to be minimal, i.e.\ a basis, with
respect to these operations. The proof establishes a connection to a
special background configuration. The number of basis invariants in
this background is shown to 
be equal to the the number of basis invariants in general backgrounds.

An additional transformation, which can be used in
some cases, is the mirror transformation. The
proof of linear independence fails in this case. 
On the other hand, by counting the invariants the proof can be
maintained up to at least 16 mass dimensions. But a general proof has
not yet been found.

An open question is how a basis for invariants in Abelian gauge
theories can be defined. Here the
commutativity of factors leads to a different structure of possible
invariants. Particularly, the division of derivatives into left, right, and
middle ones is not reasonable in this case, at least in the sense
described in Section 3. Perhaps one can find another way of
distinguishing three kinds of derivatives to apply the Bianchi identity
appropriately.

\section*{Acknowledgments} 
 
The author would like to thank Christian Schubert for drawing his 
attention to the problem treated in this paper and useful remarks
about the manuscript. The author would also like to thank Denny
Fliegner for various discussions on the reduction algorithm, Stephan
Riemersma for his helpful comments about the manuscript, and A.\ V.\
Lanyov for discussions at the AIHENP-95 workshop.

\begin{appendix}
\section{Formulae for handling fully symmetrized derivatives}
First we derive as generalization of Equation (\ref{commute}) a
commutation rule
\begin{eqnarray}\label{general commute}
  \comm{D_\mu}{D_{(\alpha_1\cdots\alpha_k)}}X&=&
  \sum_{\kappa=1}^kD_{(\alpha_1\cdots\alpha_{\kappa-1}|}
  \comm{D_\mu}{D_{|\alpha_\kappa}}
  D_{\alpha_{\kappa+1}\cdots\alpha_k)}X\nonumber\\
  &=&-\i\sum_{\kappa=1}^kD_{(\alpha_1\cdots\alpha_{\kappa-1}|}
  \comm{F_{\mu|\alpha_\kappa}}
  {D_{\alpha_{\kappa+1}\cdots\alpha_k)}X}\nonumber\\
  &=&-\i\sum_{\kappa=1}^k\sum_{\lambda=1}^\kappa
  \binomial{\kappa-1}{\lambda-1}
  \comm{D_{(\alpha_1\cdots\alpha_{\lambda-1}|}F_{\mu|\alpha_\lambda}}
  {D_{\alpha_{\lambda+1}\cdots\alpha_k)}X}\nonumber\\
  &=&-\i\sum_{\lambda=1}^k
  \binomial{k}{\lambda}
  \comm{D_{(\alpha_1\cdots\alpha_{\lambda-1}|}F_{\mu|\alpha_\lambda}}
  {D_{\alpha_{\lambda+1}\cdots\alpha_k)}X}.
\end{eqnarray}
In the first line we represent the composed commutator by elementary
commutators, then we use Equation (\ref{commute}). In the second line,
we use Leibniz' product rule for multiple symmetrized
derivatives. Finally, the order of the two summations is changed and
the inner sum is performed.

Equation (\ref{general commute}) helps to deduce Equation (\ref{symm
  recursion}):
\begin{eqnarray}
  D_\mu D_{(\alpha_1\cdots\alpha_k)}X&=&
  D_{(\mu\alpha_1\cdots\alpha_k)}X+
  \frac{1}{k+1}\sum_{\kappa=1}^k
  \comm{D_\mu}{D_{(\alpha_1\cdots\alpha_\kappa|}}
  D_{|\alpha_{\kappa+1}\cdots\alpha_k)}X\nonumber\\
  &=&D_{(\mu\alpha_1\cdots\alpha_k)}X-
  \frac{\i}{k+1}\sum_{\kappa=1}^k\sum_{\lambda=1}^{\kappa}
  \binomial{\kappa}{\lambda}
  \comm{D_{(\alpha_1\cdots\alpha_{\lambda-1}|}F_{\mu|\alpha_\lambda}}
  {D_{\alpha_{\lambda+1}\cdots\alpha_k)}X}\nonumber\\
  &=&D_{(\mu\alpha_1\cdots\alpha_k)}X-
  \frac{\i}{k+1}\sum_{\lambda=1}^k
  \binomial{k+1}{\lambda+1}
  \comm{D_{(\alpha_1\cdots\alpha_{\lambda-1}|}F_{\mu|\alpha_\lambda}}
  {D_{\alpha_{\lambda+1}\cdots\alpha_k)}X}.\nonumber
\end{eqnarray}
The first line symmetrizes the derivatives on the left-hand side by
commutators. These are replaced by means of Equation (\ref{general
  commute}). Finally, the two summations are interchanged and the
inner sum is executed.

The reversal of (\ref{symm recursion}) is
useful if a single derivative must be extracted from a symmetrized
derivative to prepare for the integration by parts. To prepare the
application of the Bianchi identity, we need a similar equation where
the single derivative stands closest to $X$. Forming the difference
between Equations (\ref{symm recursion}) and 
(\ref{general commute}) we obtain
\begin{equation}
  D_{(\alpha_1\cdots\alpha_k)}D_\mu X=
  D_{(\alpha_1\cdots\alpha_k\mu)}X+
  \frac{\i}{k+1}\sum_{\lambda=1}^k
  \lambda\binomial{k+1}{\lambda+1}
  \comm{D_{(\alpha_1\cdots\alpha_{\lambda-1}|}F_{\mu|\alpha_\lambda}}
  {D_{\alpha_{\lambda+1}\cdots\alpha_k)}X}.
\end{equation}
By means of this identity we can generalize the Bianchi identity to
\begin{eqnarray}
  \lefteqn{
    D_{(\alpha_1\cdots\alpha_k\mu)}F_{\kappa\lambda}+
    D_{(\alpha_1\cdots\alpha_k\kappa)}F_{\lambda\mu}+
    D_{(\alpha_1\cdots\alpha_k\lambda)}F_{\mu\kappa}=}
  \hspace{10ex}\nonumber\\
  &&-\frac{\i}{k+1}\sum_{\nu=1}^k\nu\binomial{k+1}{\nu+1}\Big(
  \comm{D_{(\alpha_1\cdots\alpha_{\nu-1}|}F_{\mu|\alpha_\nu}}
  {D_{\alpha_{\nu+1}\cdots\alpha_k)}F_{\kappa\lambda}}+\nonumber\\[-1.3ex]
  &&\hspace{22.9ex}
  +\comm{D_{(\alpha_1\cdots\alpha_{\nu-1}|}F_{\kappa|\alpha_\nu}}
  {D_{\alpha_{\nu+1}\cdots\alpha_k)}F_{\lambda\mu}}+\nonumber\\
  &&\hspace{22.9ex}
  +\comm{D_{(\alpha_1\cdots\alpha_{\nu-1}|}F_{\lambda|\alpha_\nu}}
  {D_{\alpha_{\nu+1}\cdots\alpha_k)}F_{\mu\kappa}}\Big).
\end{eqnarray}

These equations allow one in principle to execute the algorithm with
symmetrized derivatives.

\section{The number of basis invariants up to mass dimension 16}

This appendix lists the numbers of independent basis invariants up to
mass dimension 16, subdivided by the number of $V$- and
$F$-factors. 
Also, the basis invariants up to mass dimension eight are
given. In the tables, the number of $F$-factors in an
invariant is denoted by $f$, the number of $V$-factors by $v$.
The numbers of invariants were obtained using a C language computer
program, which will soon be made available
as described on page \pageref{homepage}.
\begin{table}[h]
\caption{\footnotesize The numbers of invariants with mass dimension six, with or without
  identification under reflection invariance. The total number of
  independent basis invariants is 5.} 

\vspace{1ex}\footnotesize
\noindent\begin{tabular}{|c||c|c|c|c||c||c|}
\hline
    $v$&\makebox[1cm]{$f=0$}&\makebox[1cm]{1}&\makebox[1cm]{2}&  
    \makebox[1cm]{3}& \makebox[1cm]{Total}&Basis invariants\\
\hline
0&   &   &  1  &  1  &  2&
\normalsize\rule{0mm}{2.2ex}$D_\mu F_{\kappa\lambda}D_\mu
F_{\kappa\lambda}|F_{\mu\nu}F_{\kappa\mu}F_{\nu\kappa}$\\
    1&   &   &  1  &   &  1&
\normalsize$VF_{\mu\nu}F_{\mu\nu}$\\
    2&     1&    &    &    &1&\normalsize$D_\mu VD_\mu V$\\
    3&     1&    &    &    &1&\normalsize$V^3$\\
\hline
Total&     2&   0&     2&     1&     5&\\
\hline
\end{tabular}
\end{table}
\vfill

\begin{table}
\caption{\footnotesize The numbers of invariants with mass dimension eight, 
{\bf with} or {\it without} identification under reflection invariance.
The total number of independent basis invariants is {\bf 17} or {\it 18},
respectively.}

\vspace{1ex}\footnotesize
\noindent\begin{tabular}{|c||c|c|rr|c|c||rr||c|}
\hline
    $v$&\makebox[1cm]{$f=0$}&\makebox[1cm]{1}&
    \multicolumn{2}{c|}{\makebox[1cm]{2}}&
    \makebox[1cm]{3}&\makebox[1cm]{4}&
    \multicolumn{2}{c||}{\makebox[1cm]{Total}}&Basis invariants\\
\hline
    0& & &\multicolumn{2}{c|}{1}&2&4&\multicolumn{2}{c||}{7}&\\
    1& & &{\bf 3}& {\it 4}&1& &{\bf 4}& {\it 5}&\\
    2&1&1&\multicolumn{2}{c|}{2}& & &\multicolumn{2}{c||}{4}&
    see Table \ref{inv8table}\\
    3&1& &\multicolumn{2}{c|}{}& & &\multicolumn{2}{c||}{1}&\\
    4&1& &\multicolumn{2}{c|}{}& & &\multicolumn{2}{c||}{1}&\\
\hline
Total&3&1&{\bf 6}&{\it 7}&3&4&{\bf 17}& {\it 18}&\\
\hline
\end{tabular}
\end{table}
\vfill

\begin{table}
\caption{\footnotesize Basis invariants with mass dimension eight,
  ordered according 
  to the numbers of $V$- and $F$-factors. The invariant in parentheses
  is generated by a mirror transformation on the precedent invariant
  in the list. Therefore it belongs to the basis only if we do not
  identify invariants by the mirror transformation.} 
\label{inv8table}

\vspace{1ex}\footnotesize
\begin{tabular}{|c|c||c|}
\hline
$v$&$f$&Basis invariants with mass dimension eight\\
\hline
0&2&\normalsize$D_{(\mu} D_{\nu)} F_{\kappa\lambda}D_{(\mu} D_{\nu)}
  F_{\kappa\lambda}$\\
0&3&
\normalsize$F_{\mu\nu}D_\mu F_{\kappa\lambda}D_\nu
F_{\kappa\lambda},F_{\mu\nu}D_\lambda F_{\kappa\mu}D_\lambda
F_{\nu\kappa}$\\ 
0&4&
\normalsize$F_{\mu\nu}F_{\mu\nu}F_{\kappa\lambda}F_{\kappa\lambda},
    F_{\mu\nu}F_{\kappa\lambda}F_{\mu\nu}F_{\kappa\lambda},
    F_{\mu\nu}F_{\kappa\mu}F_{\lambda\kappa}F_{\nu\lambda},
    F_{\mu\nu}F_{\kappa\mu}F_{\lambda\nu}F_{\kappa\lambda}$\\
1&2&\normalsize$VD_\kappa F_{\mu\nu}D_\kappa F_{\mu\nu},
D_\kappa VD_\kappa F_{\mu\nu}F_{\mu\nu},
(D_\kappa VF_{\mu\nu}D_\kappa F_{\mu\nu}),
D_{(\kappa}D_{\lambda)}VF_{\mu\kappa}F_{\lambda\mu}$\\
1&3&\normalsize$VF_{\mu\nu}F_{\kappa\mu}F_{\nu\kappa}$\\
2&0&\normalsize$D_{(\kappa}D_{\lambda)}VD_{(\kappa}D_{\lambda)}V$\\
2&1&\normalsize$D_\kappa VD_\lambda VF_{\kappa\lambda}$\\
2&2&\normalsize$V^2F_{\mu\nu}F_{\mu\nu},VF_{\mu\nu}VF_{\mu\nu}$\\
3&0&\normalsize$VD_\kappa VD_\kappa V$\\
4&0&\normalsize$V^4$\\
\hline
\end{tabular}
\end{table}

\begin{table}
\caption{\footnotesize The numbers of invariants with mass dimension ten,
{\bf with} and {\it without} identification under reflection invariance.
The total number of independent basis invariants is {\bf 79} or {\it 105},
respectively.}

\vspace{1ex}\footnotesize
\noindent\begin{tabular}{|c||rr|rr|rr|rr|rr|c||rr|}
\hline
$v$&\multicolumn{2}{c|}{$f=0$}&
\multicolumn{2}{c|}{\makebox[1cm]{1}}&\multicolumn{2}{c|}{2}&
\multicolumn{2}{c|}{3}&\multicolumn{2}{c|}{4}&
\makebox[1cm]{5}&
\multicolumn{2}{c|}{Total}\\
\hline
0&\multicolumn{2}{c|}{}&\multicolumn{2}{c|}{}&\multicolumn{2}{c|}{1}&
{\bf 4}&{\it 5}& {\bf 18}&{\it 24}&
{6}& {\bf 29}&{\it 36}\\ 
1&\multicolumn{2}{c|}{}&\multicolumn{2}{c|}{}&  
{\bf 6}&{\it 9}& {\bf 12}&{\it 21}&  
\multicolumn{2}{c|}{6}&& {\bf 24}&{\it 36}\\
2&\multicolumn{2}{c|}{1}& {\bf 2}&{\it 3}& {\bf 12}&{\it 17}&
\multicolumn{2}{c|}{2}&\multicolumn{2}{c|}{}&&
{\bf 17}&{\it 23}\\
3&\multicolumn{2}{c|}{2}&{\bf 2}&{\it 3}&\multicolumn{2}{c|}{2}&
\multicolumn{2}{c|}{}&\multicolumn{2}{c|}{}&&
{\bf 6}&{\it 7}\\
4&\multicolumn{2}{c|}{2}&\multicolumn{2}{c|}{}&\multicolumn{2}{c|}{}&
\multicolumn{2}{c|}{}&\multicolumn{2}{c|}{}&&
\multicolumn{2}{c|}{2}\\
5&\multicolumn{2}{c|}{1}&\multicolumn{2}{c|}{}&\multicolumn{2}{c|}{}&
\multicolumn{2}{c|}{}&\multicolumn{2}{c|}{}&&
\multicolumn{2}{c|}{1}\\
\hline
Total&\multicolumn{2}{c|}{6}&{\bf 4}&{\it 6}& {\bf 21}&{\it 29}& 
{\bf 18}&{\it 28}& {\bf 24}&{\it 30}&{6}& 
{\bf 79}&{\it 105}\\
\hline
\end{tabular}
\end{table}

\begin{table}
\caption{\footnotesize The numbers of invariants with mass dimension twelve,
with identification under reflection invariance.
The total number of independent basis invariants is 554.}

\vspace{1ex}\footnotesize
\begin{tabular}{|c||r|r|r|r|r|r|r||r|}
\hline
$v$&\makebox[1cm][r]{$f=0$}&\makebox[1cm][r]{1}&\makebox[1cm][r]{2}&
\makebox[1cm][r]{3}&\makebox[1cm][r]{4}&\makebox[1cm][r]{5}& 
\makebox[1cm][r]{6}&\makebox[1cm]{Total}\\
\hline
    0&&& 1& 7&68&92&28 &196\\
    1&&&10&58&102&14& & 184\\
    2& 1& 4&46&52&16&& & 119\\
    3& 3& 9&24& 3&&&&39\\
    4& 6& 4& 3&&&&&13\\
    5& 2&&&&&&& 2\\
    6& 1&&&&&&& 1\\
\hline
Total&13&17&84 & 120 & 186 & 106&28 & 554\\
\hline
\end{tabular}
\end{table}

\begin{table}
\caption{\footnotesize The numbers of invariants with mass dimension twelve, without identification
  under reflection invariance. The
total number of independent basis invariants: 902.}

\vspace{1ex}\footnotesize
\noindent\begin{tabular}{|c||r|r|r|r|r|r|r||r|}
\hline
$v$&\makebox[1cm][r]{$f=0$}& \makebox[1cm][r]{1}& \makebox[1cm][r]{2}&
\makebox[1cm][r]{3}& \makebox[1cm][r]{4}& \makebox[1cm][r]{5}&
\makebox[1cm][r]{6} & \makebox[1cm][r]{Total}\\
\hline
    0& & &  1& 10&107&151& 31&300\\
    1& & & 16&111&180& 22& &329\\
    2&  1&  6& 74& 92& 18& & &191\\
    3&  4& 17& 38&  4& & & & 63\\
    4&  7&  6&  3& & & & & 16\\
    5&  2& & & & & & &  2\\
    6&  1& & & & & & &  1\\
\hline
Total& 15& 29&132&217&305&173& 31&902\\
\hline
\end{tabular}
\end{table}

\begin{table}
\caption{\footnotesize The numbers of invariants with mass dimension 14,
with identification under reflection invariance. The
total number of independent basis invariants is 5283.}

\vspace{1ex}\footnotesize
\noindent\begin{tabular}{|c||r|r|r|r|r|r|r|r||r|}
\hline
$v$&\makebox[1cm][r]{$f=0$}& \makebox[1cm][r]{1}& \makebox[1cm][r]{2}&
\makebox[1cm][r]{3}& \makebox[1cm][r]{4}& \makebox[1cm][r]{5}&
\makebox[1cm][r]{6} &\makebox[1cm][r]{7}& \makebox[1cm][r]{Total}\\
\hline
    0& & &  1& 10&188&742&760& 87&1788\\
    1& & & 15&189&866&758& 83& &1911\\
    2&  1&  6&130&456&461& 42& & &1096\\
    3&  4& 30&170&142& 24& & & &370\\
    4& 12& 30& 50&  4& & & & & 96\\
    5&  9&  6&  3& & & & & & 18\\
    6&  3& & & & & & & &  3\\
    7&  1& & & & & & & &  1\\
\hline
Total& 30& 72&369&801&1539&1542&843& 87&5283\\
\hline
\end{tabular}
\end{table}

\begin{table}
\caption{\footnotesize The numbers of invariants with mass dimension 14, 
without identification under reflection invariance. The
total number of independent basis invariants is 9749.}

\vspace{1ex}\footnotesize
\noindent\begin{tabular}{|c||r|r|r|r|r|r|r|r||r|}
\hline
$v$&\makebox[1cm][r]{$f=0$}& \makebox[1cm][r]{1}& \makebox[1cm][r]{2}&
\makebox[1cm][r]{3}& \makebox[1cm][r]{4}& \makebox[1cm][r]{5}&
\makebox[1cm][r]{6} &\makebox[1cm][r]{7}& \makebox[1cm][r]{Total}\\
\hline
    0& & &  1&  15& 324&1394&1364&120&3218\\
    1& & & 25& 369&1656&1475& 130& &3655\\
    2&  1& 10&227& 878& 838&  66&  & &2020\\
    3&  5& 57&308& 270&  30&  &  & & 670\\
    4& 16& 56& 81&   5&  &  &  & & 158\\
    5& 11& 10&  3&  &  &  &  & &  24\\
    6&  3& & &  &  &  &  & &   3\\
    7&  1& & &  &  &  &  & &   1\\
\hline
Total& 37&133&645&1537&2848&2935&1494&120&9749\\
\hline
\end{tabular}
\end{table}

\begin{table}
\caption{\footnotesize The numbers of invariants with mass dimension 16,
with identification under reflection invariance. The
total number of independent basis invariants is 65346.}

\vspace{1ex}\footnotesize
\noindent\begin{tabular}{|c||r|r|r|r|r|r|r|r|r||r|}
\hline
$v$&\makebox[1cm][r]{$f=0$}& \makebox[1cm][r]{1}& \makebox[1cm][r]{2}&
\makebox[1cm][r]{3}& \makebox[1cm][r]{4}& \makebox[1cm][r]{5}&
\makebox[1cm][r]{6} &\makebox[1cm][r]{7}&\makebox[1cm][r]{8}&
\makebox[1cm][r]{Total}\\
\hline
    0& & &  1& 14&449&  3886& 10649&  6471&524& 21994\\
    1& & & 21&483&  4590& 11706&  7014&438& & 24252\\
    2&  1&  9&311&  2448&  6510&  3762&292& & & 13333\\
    3&  5& 75&826&  2092&  1368& 86& & & &  4452\\
    4& 25&152&544&330& 44& & & & &  1095\\
    5& 28& 72& 81&  5& & & & & &186\\
    6& 17&  9&  4& & & & & & & 30\\
    7&  3& & & & & & & & &  3\\
    8&  1& & & & & & & & &  1\\
\hline
Total& 80&317&  1788&  5372& 12961& 19440& 17955&  6909&524& 65346\\
\hline
\end{tabular}
\end{table}

\begin{table}
\caption{\footnotesize The numbers of invariants with mass dimension 16,
without identification under reflection invariance. The
total number of independent basis invariants is 127072.}

\vspace{1ex}\footnotesize
\noindent\begin{tabular}{|c||r|r|r|r|r|r|r|r|r||r|}
\hline
$v$&\makebox[1cm][r]{$f=0$}& \makebox[1cm][r]{1}& \makebox[1cm][r]{2}&
\makebox[1cm][r]{3}& \makebox[1cm][r]{4}& \makebox[1cm][r]{5}&
\makebox[1cm][r]{6} &\makebox[1cm][r]{7}&\makebox[1cm][r]{8}&
\makebox[1cm][r]{Total}\\
\hline
    0& & &  1& 22&809&  7550& 20659& 12471&823& 42335\\
    1& & & 36&954&  9012& 23280& 13740&822& & 47844\\
    2&  1& 15&564&  4808& 12659&  7341&473& & & 25861\\
    3&  7&147&  1580&  4140&  2610&154& & & &  8638\\
    4& 36&291&  1005&630& 58& & & & &  2020\\
    5& 44&140&138&  7& & & & & &329\\
    6& 22& 15&  4& & & & & & & 41\\
    7&  3& & & & & & & & &  3\\
    8&  1& & & & & & & & &  1\\
\hline
Total&114&608&  3328& 10561& 25148& 38325& 34872& 13293&823&127072\\
\hline
\end{tabular}
\end{table}

\section{Representation of primitive invariants by general invariants
  in the special background Eqs. (\ref{special AV})}
\label{appendix rules}

We start with an arbitrary primitive invariant. Products $A_\mu A_\mu$
are replaced by $-V$ at once. All contractions between $A$'s are
removed by
\begin{equation}\label{rep1}
  \ldots A_\mu XA_\mu\ldots=
  \frac{1}{2}\ldots\left(-VX-XV+D^2X\right)\ldots\quad.
\end{equation}
$X$ is a product of $A$'s and/or factors built from $D$, $F$, and/or
$V$. These factors are expressed by simple factors as described in
Section 2. $D^2X$ is transformed by the product rule and by
$D_\mu A_\nu = -\i\comm{A_\mu}{A_\nu}=F_{\mu\nu}$ into expressions
containing no derivatives of $A$.

We are left with expressions where all $A$'s are
contracted with simple factors. These $A$'s are removed one after the
other. First, they are moved next to the associated factors by
\begin{equation}\label{rep2}
  A_\mu XY_\mu=\comm{A_\mu}{X}Y_\mu+XA_\mu Y_\mu=
  \i D_\mu XY_\mu+XA_\mu Y_\mu
\end{equation}
$D_\mu X$ is treated like $D^2X$ before. We have to find an expression
for $A_\mu Y_\mu$. If the index $\mu$ in $Y_\mu$ belongs to a
derivative we commute it to become the innermost derivative. If
$Y_\mu$ is an 
$F$-factor we apply the Bianchi identity to ensure $\mu$ is an
index of the $F$. If then $Y_\mu$ is a derivative of something, say
$X_\mu$, we use 
\begin{equation}\label{rep3}
  A_\mu D_\rho X_\mu=D_\rho\left(A_\mu X_\mu\right)-F_{\rho\mu}X_\mu,
\end{equation}
so we have to find an expression for $A_\mu X_\mu$. Repeating the last
step we arrive at $A_\mu F_{\mu\nu}$ or $A_\mu D_\mu V$ which are
expressed as
\begin{equation}\label{rep4}
  A_\mu F_{\mu\nu}=
  \frac{1}{2}D_\nu V+\frac{\i}{2}D_\mu F_{\mu\nu},\qquad
  A_\mu D_\mu V=\frac{\i}{2}D^2V.
\end{equation}
Having removed in this way all $A$'s contracted with simple factors,
we end up with a representation of the primitive invariant completely
in terms of $D$, $F$, and $V$.
\end{appendix}


\begin{thebibliography}{xx}

\bibitem{effac}
E.g. C. Itzykson, J.-B. Zuber,
{\bibit Quantum Field Theory} (McGraw-Hill, New York, 1980);\\
R. Jackiw,
{\bibit Phys. Rev.} {\bibbf D9}, 1686 (1974).

\bibitem{SDW}
J. Schwinger, {\bibit Phys. Rev.} {\bibbf 82}, 664 (1951);\\
B. DeWitt, {\bibit Dynamical Theory of Groups and Fields} (Gordon and
Breach, New York, 1965).

\bibitem{GS}
R. T. Seeley, {\bibit Amer. Math. Soc.} {\bibbf 10}, 288 (1967);\\
P. B. Gilkey, {\bibit J. Diff. Geom.} {\bibbf 10}, 601 (1975).

\bibitem{H}
J. Hadamard, {\bibit Lectures on Cauchy's Problem in Linear Partial
  Differential Equations} (Yale Univ. Press, New Haven, 1923).

\bibitem{belkov etal}
A. A. Bel'kov, A.~V.~Lanyov, A. Schaale, 
{\bibit Comput. Phys. Commun.} {\bibbf 95}, 123 (1996).

\bibitem{worldline formalism}
M. J. Strassler,
{\bibit Nucl. Phys.} {\bibbf B385}, 145 (1992);\\
M. G. Schmidt, C. Schubert,
{\bibit Phys. Lett.} {\bibbf B318}, 438 (1993);\\
D.~Fliegner, M.~G.~Schmidt, C.~Schubert,
{\bibit Z.~Phys.} {\bibbf C64}, 111 (1994);\\
D.~Fliegner, P.~Haberl, M.~G.~Schmidt, C.~Schubert,
{\bibit Discourses in Mathematics and its Applications} No. {\bibbf
  4}, 87 (Texas A\&M 1995), {\tt hep-th/9411177};
in {\bibit New Computing Techniques in Physics Research IV,
Proceedings of the AIHENP-95 workshop, Pisa, 1995}, 
eds. B. Denby, D. Perret-Gallix (World Scientific, Singapore, 1995),
p. 199, {\tt hep-th/9505077}.  

\bibitem{avramidi}
I.~G.~Avramidi,
{\bibit Nucl.~Phys.} {\bibbf B355}, 712 (1991);\\
I. G. Avramidi, R. Schimming,
{\bibit Quantum Field Theory Under the Influence of External
  Conditions}, ed. M. Bordag (Teubner, Stuttgart, 1996), p. 150. 

\bibitem{FKWC}
S.~A.~Fulling, R.~C.~King, B.~G.~Wybourne, C.~J.~Cummins, 
{\bibit Class.~Quantum Grav.} {\bibbf 9}, 1151 (1992),
and references therein.

\bibitem{van de Ven}
A.~E.~M.~van~de~Ven,
{\bibit Nucl. Phys.} {\bibbf B250}, 593 (1985).

\bibitem{bv}
A. O. Barvinsky, G. A. Vilkovisky,
{\bibit Nucl.~Phys.} {\bibbf B282}, 163 (1987);
{\bibit ibid.} {\bibbf B333}, 471 (1990);
{\bibit ibid.} {\bibbf B333}, 512 (1990).

\bibitem{BGV}
A.~O.~Barvinsky, Yu.~V.~Gusev, G.~A.~Vilkovisky, V.~V.~Zhytnikov,
{\bibit Nucl.~Phys.} {\bibbf B439}, 561 (1995);
Preprint 93-0274, Univ.~of Manitoba, Winnipeg, February 1993;
{\bibit J.~Math.~Phys.} {\bibbf 35}, 3525 (1994).

\bibitem{proceedings}
U. M\"uller, 
in {\bibit New Computing Techniques in Physics Research IV,
Proceedings of the AIHENP-95 workshop, Pisa, 1995}, 
eds. B. Denby, D. Perret-Gallix (World Scientific, Singapore, 1995),
p. 19, {\tt hep-th/9508031}.

\bibitem{dimreg}
G.~'t Hooft, M. Veltman,
{\bibit Nucl. Phys.} {\bibbf B44},189 (1972);\\
C. G. Bollini, J. J. Giambiagi, 
{\bibit Nuovo Cim.} {\bibbf 12B}, 20 (1972);\\
J. F. Ashmore,
{\bibit Nuovo Cim. Lett.} {\bibbf 4}, 289 (1972).

\bibitem{zetareg}
S. W. Hawking,
{\bibit Commun. Math. Phys.} {\bibbf 55}, 133 (1977).

\bibitem{methods}
G.~'t Hooft,
{\bibit Nucl.~Phys.} {\bibbf B62}, 444 (1973);\\
K.~Kikkawa,
{\bibit Prog.~Theor.~Phys.} {\bibbf 56}, 947 (1976);\\
E.~Onofri,
{\bibit Am.~J.~Phys.} {\bibbf 46}, 379 (1978);\\
Y.~Fujiwara, T.~A.~Osborn, S.~F.~J.~Wilk,
{\bibit Phys.~Rev.} {\bibbf A25}, 14 (1982);\\
R.~MacKenzie, F.~Wilczek, A.~Zee,
{\bibit Phys.~Rev.~Lett.} {\bibbf 53}, 2203 (1984);\\
C.~M.~Fraser,
{\bibit Z.~Phys.} {\bibbf C28}, 101 (1985);\\
R.~I.~Nepomechie,
{\bibit Phys.~Rev.} {\bibbf D31}, 3291 (1985);\\
J.~A.~Zuk,
{\bibit J.~Phys.} {\bibbf A18}, 1795 (1985);
{\bibit Phys.~Rev.} {\bibbf A33}, 4342 (1986);
{\bibit ibid.} {\bibbf D34}, 1791 (1986);
{\bibit Nucl.~Phys.} {\bibbf B280}, 125 (1987);\\
A. O. Barvinsky, T. A. Osborn, Yu. V. Gusev,
{\bibit J. Math. Phys.} {\bibbf 36}, 30 (1995).

\bibitem{results}
P.~Amsterdamski, A.~L.~Berkin, D.~J.~Connor,
{\bibit Class.~Quantum Grav.} {\bibbf 6}, 1981 (1989);\\
L.~Carson,
{\bibit Phys.~Rev.} {\bibbf D42}, 2853 (1990);\\
A.~A.~Belkov, D.~Ebert, A.~V.~Lanyov, A.~Schaale,
{\bibit Int.~J.~Mod.~Phys.} {\bibbf C4}, 775 (1993);
{\bibit ibid.} {\bibbf A8}, 1313 (1993);\\
J.~Caro, L.~L.~Salcedo,
{\bibit Phys.~Lett.} {\bibbf B309}, 359 (1993);\\
A.~O.~Barvinsky, Yu.~V.~Gusev,
MANIT-11-94, Univ.~of Manitoba, Winnipeg, 1994.

\end{thebibliography}
\end{document}